\documentclass[onecolumn,english]{revtex4-1}
\usepackage[T1]{fontenc}
\usepackage[latin9]{luainputenc}
\usepackage{verbatim}
\usepackage{amsmath}
\usepackage{amssymb}
\usepackage{graphicx}
\usepackage{babel}

\begin{document}
\title{Noncommutative phase-space effects in thermal diffusion of Gaussian states}
\author{Jonas F. G. Santos}
\affiliation{Centro de Ciências Naturais e Humanas, Universidade Federal do ABC,
Avenida dos Estados 5001, 09210-580 Santo André, São Paulo, Brazil}
\email{jonas.floriano@ufabc.edu.br}

\begin{abstract}
Noncommutative phase-space and its effects have been studied in different settings in physics, in order to unveil a better understanding of phase-space structures. Here, we use the thermal diffusion approach to study how noncommutative effects can influence the time evolution of a one-mode Gaussian state when in contact with a thermal environment obeying the Markov approximation. Employing the cooling process and considering the system of interest as a one-mode Gaussian state, we show that the fidelity comparing the Gaussian state of the system in different times and the asymptotic thermal state is useful to sign noncommuative effects. Besides, by using the monotonicity behavior of the fidelity, we discuss some aspects of non-Markovianity during the dynamics.
\end{abstract}

\maketitle


\section{Introduction}

Quantum dissipative systems is a research field broadly investigated
in theoretical and experimental physics because they are one of the
best examples of open quantum system applications \cite{Petruccione01,Rivas01}.
There is a fairly deep interest in dissipative acting on quantum systems,
ranging from quantum optics to condensed matter physics as, for instance,
entangled states in optical cavities \cite{Reiter}, dissipative optomechanical
systems \cite{Liew}, in the Rabi model \cite{Plenio01}, and in systems
interacting with Gaussian dissipative reservoir \cite{Marcin}. Besides,
there exist a practical appeal to control dissipation in quantum technologies,
such as quantum information processing \cite{Pleni02} and quantum
computation \cite{Zureck01}. Then, in general, one can think that
dissipation comes from the interaction of quantum systems to some
external agent, such as a thermal environment.

On the other hand, from a theoretical point of view, exciting questions
arise when new quantum features can affect quantum thermodynamic processes,
in particular, that features underlying the context of phase-space
noncommutativity (NC) extension of quantum mechanics. The noncommutativity
in the configuration space has been firstly suggested by Snyder \cite{Snyder}
as a propose to avoid divergences in the quantum field theory. Additionally,
there is a fairly deep consensus that in the Planck scale $(\sim10^{-32}cm)$,
the notion of space-time has to be drastically rectified in a consistent
formulation of quantum mechanics and gravity \cite{ref01,ref02},
such that noncommutativity must be assumed at high energy scales.
There is a large number of studies concerning the implications of
what has been conventionally called by noncommutative quantum mechanics
(NCQM), for instance, in the context of $2D$-harmonic oscillator
\cite{Vergara,Bernardini01}, the gravitational quantum well \cite{Orfeu01,Banerjee, extra01},
and in relativist dispersion relations \cite{Leal01}.

In the context of experimental verification of signatures of noncommutative effects, it is worth to mention that current experimental range is far away from the Planck scale and, at this moment is not clear how to reach this experimental ability. However, some recent theoretical developments in this direction have arisen, for instance, using quantum optics \cite{extra02} and opto-mechanical \cite{extra03} setups. In particular, Ref.  \cite{extra02} employs a quantum mechanical ancillary system that can work in order to measure any deformation of the canonical commutator directly, by means of an additional optical phase that is absent in the standard quantum mechanics. Moreover, some theoretical studies have been done exploring noncommutative effects in master equation, in particular for the Brownian particle \cite{extra06}. In the direction of these recent developments, we could argue that some tools from information theory would be suitable to sign noncommutative effects in open quantum systems. For instance, the quantum fidelity is a good measure to indicate any non-trivial signature in the dynamics of a system evolving in thermal contact with an environment.

It can be observed that for systems described by Hamiltonians at most
quadratic in their coordinates, the noncommutative effects can be
effectively mapped to the standard quantum mechanics as an external
magnetic field acting on the system \cite{Bernardini01,Jonas00,Jonas01,Jonas02}.
This fact allows to obtain a convenient correspondence between the NCQM
and the standard quantum mechanics (SQM) which is suitable to investigate how NC effects could impact
some particular dynamics. The aim of this work is to use this property
to study how NC effects can influence a thermalization process. We
considered the so called Gaussian states as our system evolving under
the dynamics of the NC harmonic oscillator. Then, we put the system
in a contact with a thermal environment in order to probe how the
time evolution of the system state is influenced by NC effects. We
assume the cooling process, i. e., the system and the thermal environment
are associated to the mean numbers of photons $\bar{n}$ and $\bar{m}$,
respectively, such that $\bar{n}>\bar{m}$. To quantify our study
we use the quantum fidelity which has a well known form when both
states are Gaussian. 

This work is organized as follows. In section \ref{sec:Theoretical-Framework}
we introduce the necessary information on noncommutative quantum mechanics
and the Seiberg-Witten map. Besides, we provide the theoretical framework
to treat thermal diffusion of Gaussian states. In particular, the
section \ref{sec:Mapping-Noncommutative-Effects} is devoted to show
how to map NC effects as effective external fields acting on the harmonic
oscillator Hamiltonian and its equations of motion. Section \ref{sec:Thermal-Diffusion-with}
is dedicated to show how NC effects can affect the cooling dynamics
of a Gaussian state when in contact with a thermal environment. Finally,
in section \ref{sec:Conclusions} we draw our conclusions and final
remarks.

\section{Theoretical Framework\label{sec:Theoretical-Framework}}

In this section we provide the theoretical framework to treat noncommutativity
in phase-space and the thermal diffusion equation for Gaussian states.

\subsection{Noncommutative quantum mechanics in phase-space }

Noncommutative quantum mechanics is based on the deformed Heisenberg-Weyl
algebra \cite{Bernardini01,Orfeu01,Orfeu02,Gamboa} and it is represented
by the commutation relations given by,
\begin{equation}
\small
\left[\hat{q}_{i},\hat{q}_{j}\right]=i\theta_{ij},\,\left[\hat{q}_{i},\hat{p}_{j}\right]=i\hbar\delta_{ij},\,\left[\hat{p}_{i},\hat{p}_{j}\right]=i\zeta_{ij},\quad i,j=1,...,d,\label{ncrelation}
\end{equation}
where $\theta_{ij}$ and $\zeta_{ij}$ are invertible antisymmetric
real constant $(d\times d)$ matrices, and one can define the matrix
$\Sigma_{ij}=\delta_{ij}+\theta_{ik}\zeta_{kj}/\hbar^{2}$, which
is also invertible if $\theta_{ik}\xi_{kj}\neq-\hbar^{2}\delta_{ij}$.
Writing $\theta_{ij}=\theta\epsilon_{ij}$ and $\zeta_{ij}=\zeta\epsilon_{ij}$,
with $\epsilon_{ii}=0$, $\epsilon_{ij}=-\epsilon_{ji}$, one can
interpret $\eta$ and $\zeta$ as being new constants in the quantum
theory, which have been extensively studied recently \cite{Bernardini01,Jonas02,Saha,Bastos,Bastos02,Andreas}.
Furthermore, there is a way of connecting the Hilbert space of the
NCQM to that of the SQM, which is represented
by the following relations,
\begin{equation}
\small
\left[\hat{Q}_{i},\hat{P}_{j}\right]=0,\,\left[\hat{Q}_{i},\hat{P}_{j}\right]=i\hbar\delta_{ij},\,\left[\hat{P}_{i},\hat{P}_{j}\right]=0,\quad i,j=1,...,d.
\end{equation}

This is implemented through the Seiberg-Witten (SW) map, given by
$\hat{q}_{i}=\nu\hat{Q}_{i}-(\theta/2\nu\hbar)\epsilon_{ij}\hat{P}_{j}$
and $\hat{p}_{i}=\mu\hat{P}_{i}+(\zeta/2\mu\hbar)\epsilon_{ij}\hat{Q}_{j}$,
where $\nu$ and $\mu$ are arbitrary parameters fulfilling the condition
$\theta\zeta=4\hbar^{2}\mu\nu(1-\mu\nu)$.

Once the SW map is applied to the Hamiltonian of a system, to describe
the state of such a system, one can use the density matrix $\hat{\rho}=|\Psi\rangle\langle\Psi|$
, which can be used to define the associated Wigner function through
the Weyl transform \cite{Case,Zachos},
\small
\begin{eqnarray*}
W(Q_{i},P_{i}) & = & h^{-1}\rho^{W}\\
 & = & h^{-1}\int dy\,exp[iPy/\hbar]\Psi(Q_{i}-y/2)\Psi(Q_{i}+y/2),
\end{eqnarray*}
\normalsize
which can be naturally generalized to a statistical mixture, with ``W'' standing for the Weyl transform. The marginal
integration of the Wigner function results in,
\begin{align}
\psi^{\star}(Q_{i})\psi(Q_{i}) & =\int dP_{i}\,W(Q_{i},P_{i})\\
\psi^{\star}(P_{i})\psi(P_{i}) & =\int dQ_{i}\,W(Q_{i},P_{i}),
\end{align}
i. e., the probability distribution for position and momentum, respectively.
The Wigner function can be used to obtain the expectation value of
an observable $\mathcal{\hat{\mathcal{O}}}$ as,
\begin{equation}
\langle\mathcal{O}\rangle=\int\int dQ_{i}dP_{i}\,W(Q_{i},P_{i})\mathcal{O}^{W}(Q_{i},P_{i}).
\end{equation}
where $\mathcal{O}^{W}(Q_{i},P_{i})$ is the Weyl transform of the
operator $\hat{\mathcal{O}}$.

\subsection{Gaussian States and Thermal Diffusion }

One important class of states with extensive theoretical and experimental
applications are the so called Gaussian states (GS) which has been
largely used in quantum information and quantum communication \cite{Adesso,Wang},
quantum metrology \cite{Adesso02}, quantum optics \cite{Walls} etc.
Gaussian states are a subset of the more general class used to treat
continuous variables (CV) systems \cite{Wang, extra05, Adesso}. An important
aspect of the GS is that they are completely characterized by their
first and second moments. Introducing a vector $\vec{R}(Q_{1},P_{1},Q_{2},P_{2})$
to collect the coordinates of a two-dimensional system, the first
moments can be rearranged in the vector $\vec{d}=(\langle Q_{1}\rangle,\langle P_{1}\rangle,\langle Q_{2}\rangle,\langle P_{2}\rangle)$.
The set of all second moments are collected in the so called covariance
matrix (CM), given by $\sigma=\sigma_{11}\oplus\sigma_{22}$ for a
two-mode Gaussian state, where,
\begin{equation}
\sigma_{ii}=\left(\begin{array}{cc}
\sigma_{Q_{i}Q_{i}} & \sigma_{P_{i}Q_{i}}\\
\sigma_{Q_{i}P_{i}} & \sigma_{P_{i}P_{i}}
\end{array}\right),\label{CM}
\end{equation}
with $\sigma_{AB}=\langle AB+BA\rangle-2\langle A\rangle\langle B\rangle$.
It can be shown that for a bona-fide two-mode Gaussian state the CM
satisfies the relation $\sigma+i\Omega\geq0$, where,
\begin{equation}
\Omega=\left(\begin{array}{cc}
0 & 1\\
-1 & 0
\end{array}\right)^{\oplus2},
\end{equation}
such that $\left[\vec{R}_{i},\vec{R}_{j}\right]=i\Omega_{ij}$ \cite{Simon}.
In particular, for GS the Wigner function assumes a gently form given
by \cite{Adesso},
\begin{equation}
W_{G}(\vec{R})=\frac{\text{exp}\left[-(1/2)(\vec{R}-\vec{d})\sigma^{-1}(\vec{R}-\vec{d})\right]}{(2\pi)^{2n}\sqrt{Det[\sigma]}},\label{wigner}
\end{equation}
where $n$ is the number of modes of the system. A special type
of GS is the thermal state. A one-mode thermal state is represented
by \cite{Adesso},
\begin{equation}
\rho^{th}(\bar{m})=\sum_{m=0}^{\infty}\frac{\bar{m}^{m}}{(\bar{m}+1)^{m+1}}|m\rangle\langle m|,\label{therm}
\end{equation}
where $\bar{m}=\left\{ exp\left[\hbar\omega/k_{B}T\right]-1\right\} ^{-1}$
is the mean number of photons in the bosonic mode, $k_{B}$ is the
Boltzmann constant, $T$ is the associated temperature, and $\{|m\rangle\}$
is the Fock basis, with first moments and CM given by $\vec{R}=(0,0)$
and $\sigma^{th}=(2\bar{m}+1)\mathbb{I}_{2\times2}$, respectively,
where $\mathbb{I}_{2\times2}$ represents a two-by-two identity matrix. 

The propagation of a general Gaussian state in a noisy and dissipative
channel where each mode is coupled with a different and uncorrelated
Markovian environment modeled by a stationary continuum of oscillators
can be described, in the interaction picture, by the following master
equation, 
\begin{equation}
\frac{d}{dt}\rho=-i\left[\hat{H},\hat{\rho}\right]+\mathcal{L}(\hat{\rho}),
\end{equation}
where the first and second term on the right-hand side represent the
unitary and dissipative part of the dynamics. It can be shown that
when the system of interest is a Gaussian state the dynamics is easily
evolved in terms of the first moments and covariance matrix \cite{Serafini,Giovanneti01},
\begin{eqnarray*}
\dot{\sigma} & = & \Gamma\sigma+\Gamma(2\bar{m}+1)\mathbb{I}_{2\times2},\\
\dot{\vec{d}} & = & -(\Gamma/2)\vec{d},
\end{eqnarray*}
with solutions,
\begin{align}
\sigma(t) & =e^{-\Gamma t}\sigma(0)+(1-e^{-\Gamma t})(2\bar{m}+1)\mathbb{I}_{2\times2},\\
\vec{d}(t) & =e^{-\Gamma t/2}\vec{d}(0),
\end{align}
where $\Gamma$ is the decay rate and $\bar{m}$ is the mean number
of photons of the thermal environment, and $\sigma(0)$ and $\vec{d}(0)$
are the initial covariance matrix and first moments of the system.
Assuming a mean number of photons to the system as being $\bar{n}$,
we remark that for the cooling process, $\bar{n}>\bar{m}$, whereas
for the heating process, $\bar{n}<\bar{m}$. Here, we adopt the cooling
process to investigate the NC effects. To conclude this section, we
are interested in the time evolution of the following two-dimensional
Gaussian Wigner function,
\begin{eqnarray}
\small
W_{G}(\vec{R}) & = & \frac{1}{\pi^{2}}\text{exp}\left[-\left((Q_{1}(t)-x_{0})^{2}-(Q_{2}(t)-y_{0})^{2}\right)\right]\nonumber \\
 & \times & \text{exp}\left[-\left((P_{1}(t)-p_{y_{0}})^{2}-(P_{2}(t)-p_{y_{0}})^{2}\right)\right],\label{www}
\end{eqnarray}
where $x_{0}$,$y_{0}$, $p_{x_{0}}$ and $p_{y_{0}}$ are arbitrary
initial parameters.

\section{Mapping Noncommutative Effects as External Fields\label{sec:Mapping-Noncommutative-Effects}}

Let us consider the Hamiltonian of a noncommutative harmonic oscillator,
\begin{equation}
\hat{H}^{NC}(\hat{q}_{i},\hat{p}_{i})=\frac{\hat{p}_{i}^{2}}{2m}+\frac{m\omega^{2}}{2}\hat{q}_{i}^{2},\label{HamilHO}
\end{equation}
 where $m$ and $\omega$ are the mass and frequency of the system,
respectively. Applying the SW map to obtain the standard version of
the eq.(\ref{HamilHO}) and then performing the Weyl transform, one
gets \cite{Vergara,Bernardini01},
\begin{equation}
H(Q_{i},P_{i})=\alpha^{2}Q_{i}^{2}+\beta^{2}P_{i}^{2}+\gamma\sum_{i,j=1}^{2}\epsilon_{ij}P_{i}Q_{j},\label{HamilHO01}
\end{equation}
where we defined,
\begin{equation}
\small
\alpha^{2}=\frac{\nu^{2}m\omega^{2}}{2}+\frac{\zeta^{2}}{8m\mu^{2}\hbar^{2}},\,\beta^{2}=\frac{\mu^{2}}{2m}+\frac{m\omega^{2}\theta^{2}}{8\nu^{2}\hbar^{2}},\,\gamma=\frac{\theta}{2\hbar}m\omega^{2}+\frac{\zeta}{2m\hbar}.\label{cts}
\end{equation}
Following ref. \cite{Bernardini01}, we use the Heisenberg equation to
obtain a set of uncoupled four equations of motions, 

\small
\begin{eqnarray}
Q_{1}(t) & = & x_{0}\cos(\Omega t)\cos(\gamma t)+y_{0}\cos(\Omega t)\sin(\gamma t)\nonumber \\
 & + & \frac{\beta}{\alpha}[p_{y_{0}}\sin(\Omega t)\sin(\gamma t)+p_{x_{0}}\sin(\Omega t)\cos(\gamma t)]\label{sol1}\\
Q_{2}(t) & = & y_{0}\cos(\Omega t)\cos(\gamma t)-x_{0}\cos(\Omega t)\sin(\gamma t)\nonumber \\
 & - & \frac{\beta}{\alpha}[p_{x_{0}}\sin(\Omega t)\sin(\gamma t)-p_{y_{0}}\sin(\Omega t)\cos(\gamma t)]\label{sol2}\\
P_{1}(t) & = & p_{x_{0}}\cos(\Omega t)\cos(\gamma t)+p_{y_{0}}\cos(\Omega t)\sin(\gamma t)\nonumber \\
 & - & \frac{\alpha}{\beta}[y\sin(\Omega t)\sin(\gamma t)+x\sin(\Omega t)\cos(\gamma t)]\label{sol3}\\
P_{2}(t) & = & p_{y_{0}}\cos(\Omega t)\cos(\gamma t)-p_{x_{0}}\cos(\Omega t)\sin(\gamma t)\nonumber\\
 & + & \frac{\alpha}{\beta}[x_{0}\sin(\Omega t)\sin(\gamma t)-y_{0}\sin(\Omega t)\cos(\gamma t)],\label{sol4}
\end{eqnarray}
\normalsize
where $\Omega=2\alpha\beta=\omega\sqrt{(2\mu\nu-1)^{2}+\xi^{2}}$
and $\xi=(1/2\hbar)\left[m\omega\theta+\zeta/(m\omega)\right]$. 

From eq.(\ref{HamilHO01}) we can note that the influence
of the noncommutative parameters $\theta$ and $\zeta$ on the dynamics
of the system is effectively a magnetic field-like term in the orthogonal
direction to the plane of the oscillator. More explicitly, from the
Hamiltonian of a harmonic oscillator in an external magnetic field
\cite{Rob}, one has,
\begin{equation}
B_{0}\sim\frac{m^{2}\omega^{2}\theta}{q\hbar}+\frac{\zeta}{q\hbar},
\end{equation}
where $q$ is the effective charge associated to the harmonic oscillator. 

\section{Thermal Diffusion with NC Effects \label{sec:Thermal-Diffusion-with}}

In order to see how NC effects can influence the thermal diffusion
process, we consider a Gaussian Wigner function as in eq.(\ref{www})
evolving onto the dynamics dictated by eqs.(\ref{sol1}-\ref{sol4}).
Assuming we are interested in the phase-space $\left(Q_{1},P_{1}\right)$
as our system, the respective Wigner function is obtained tracing
out the coordinates $\left(Q_{2},P_{2}\right)$, i. e., 
\begin{equation}
W_{sys}(Q_{1},P_{1})=\int_{-\infty}^{\infty}\int_{-\infty}^{\infty}dQ_{2}dP_{2\,}W_{G}(\vec{R})=W_{1}\label{wsys}
\end{equation}

It is important to stress that $W_{sys}(Q_{1},P_{1})$ will be Gaussian
during all the time evolution, such that we can assume that the first
moments and the covariance matrix are sufficient to characterize the
state of the system and they are given by $\langle\vec{d}_{sys}\rangle$
and $\sigma_{sys}$, i. e.,
\begin{equation}
\small
\langle\vec{d}_{sys}\rangle=(\langle Q_{1}\rangle_{W_{1}},\langle P_{1}\rangle_{W_{1}}),\,\sigma_{sys}=\left(\begin{array}{cc}
\langle\sigma_{Q_{1}Q_{1}}\rangle_{W_{1}} & \langle\sigma_{P_{1}Q_{1}}\rangle_{W_{1}}\\
\langle\sigma_{Q_{1}P_{1}}\rangle_{W_{1}} & \langle\sigma_{P_{1}P_{1}}\rangle_{W_{1}}
\end{array}\right),\label{sys01}
\end{equation}
where $\langle\sigma_{AB}\rangle_{W_{1}}=\langle AB+BA\rangle_{W_{1}}-2\langle A\rangle_{W_{1}}\langle B\rangle_{W_{1}}$
and the mean values are obtained over the Wigner function $W_{1}$.

After that, the system is placed in thermal contact with an environment
represented by a mean number of photons $\bar{m}$, such that the
first moments are null and $\sigma^{th}(\bar{m})=(2\bar{m}+1)\mathbb{I}_{2\times2}$.
For an environment fulfilling the Born-Markovian approximation, we
assume a decay rate $\Gamma$. Besides, we consider $\langle\hat{B}_{i}^{\dagger}(0)\hat{B}_{i}(\nu)\rangle=\bar{m}\delta(\nu)$
and $\langle\hat{B}_{i}(0)\hat{B}_{i}(\nu)\rangle=M\delta(\nu)$,
where $\hat{B}$ is the bosonic operator associated to the environment
and the second expression relates the squeezing feature of the environment.
Here, we assume that $M=0$, i. e., the environment is just a thermal
one. Furthermore, we consider that the thermal environment interacts
only with the sub-space $(Q_{1},P_{1})$ or, in other words, the interaction
between the sub-space $(Q_{2},P_{2})$ and the thermal environment
is sufficiently week such that it does not cause any effects on the
evolution of the state $W_{sys}(Q_{1},P_{1}).$ During the time evolution
of the system in contact with the thermal environment, we assume the
cooling process, i. e., we indexed to the system a mean number of
photons $\bar{n}$, with $\bar{n}>\bar{m}$ and, for a Markovian evolution,
$\bar{n}\rightarrow\bar{m}$ for a time sufficiently large. The time
evolution of the first moments and CM during the thermal diffusion
reads,
\begin{align}
\sigma_{sys}(t) & =\bar{n}e^{-\Gamma t}\sigma_{sys}(0)+(1-e^{-\Gamma t})(2\bar{m}+1)\mathbb{I}_{2\times2},\\
\vec{d}_{sys}(t) & =e^{-\Gamma t/2}\vec{d}_{sys}(0).
\end{align}
In order to quantify how NC effects mapped effectively as external
fields influence the cooling process, we use the quantum fidelity
which, for two Gaussian states, has the well known expression \cite{Holevo,Scutaru, extra04},
\begin{equation}
F(\sigma_{1},\vec{d}_{1};\sigma_{2},\vec{d}_{2})=\frac{2}{\sqrt{\Delta+\delta}-\sqrt{\delta}}e^{-\frac{1}{2}\vec{d}^{T}\sigma_{+}^{-1}\vec{d}},\label{fidelity}
\end{equation}
where $\Delta\equiv Det[\sigma_{1}+\sigma_{2}]$, $\delta=(Det[\sigma_{1}]-1)(Det[\sigma_{2}]-1)$,
$\vec{d}\equiv\vec{d}_{1}-\vec{d}_{2}$, and $\sigma_{+}=\sigma_{1}+\sigma_{2}$.
The fidelity is bounded by $0\leq F\leq1$, with $F=0$ for two completely
different states and $F=1$ for two identical states. Here we consider
$(\sigma_{1},\vec{d}_{1})$ and $(\sigma_{2},\vec{d}_{2})$ as being
our system of interest during the time evolution and the asymptotic thermal state for complete cooling, respectively.
\begin{figure}
\includegraphics[scale=0.5]{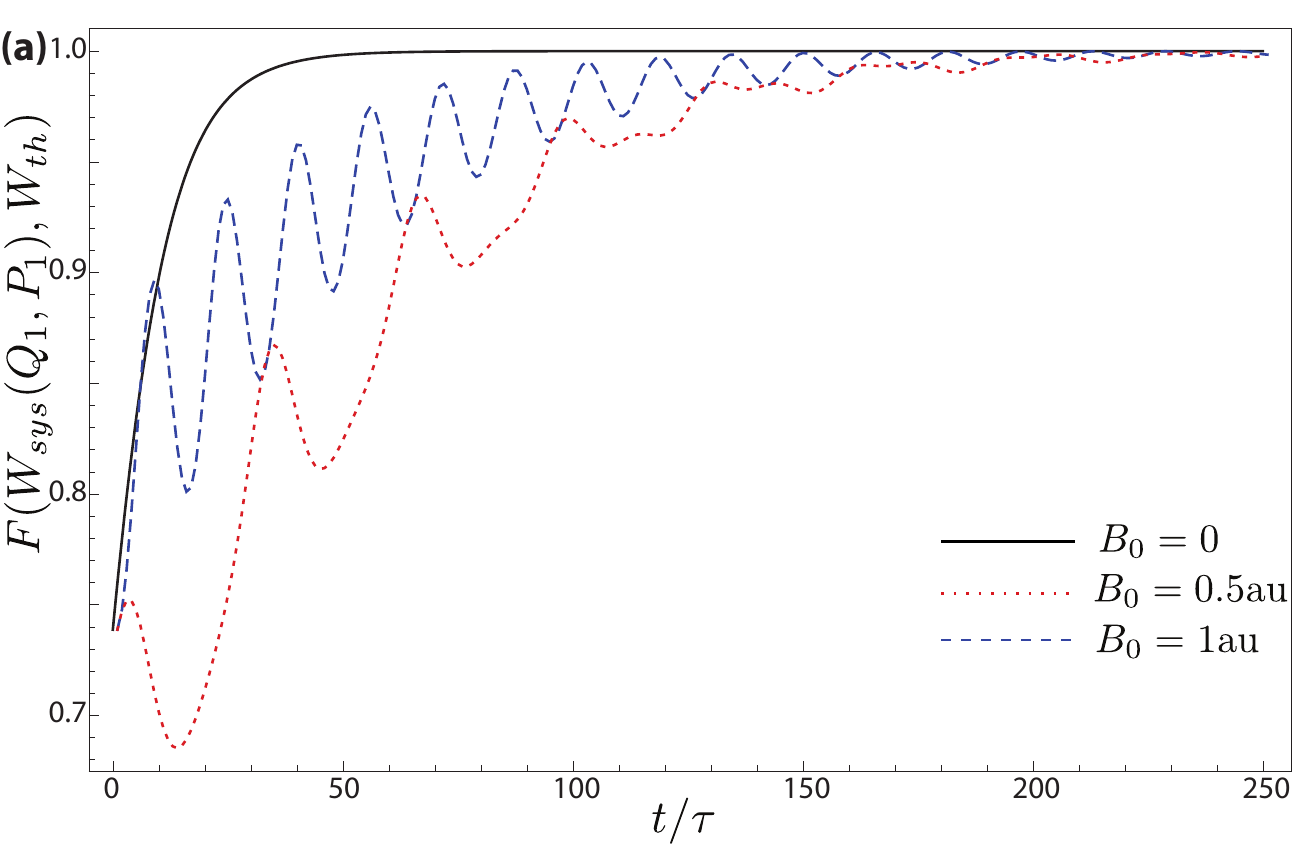}
\\
\includegraphics[scale=0.5]{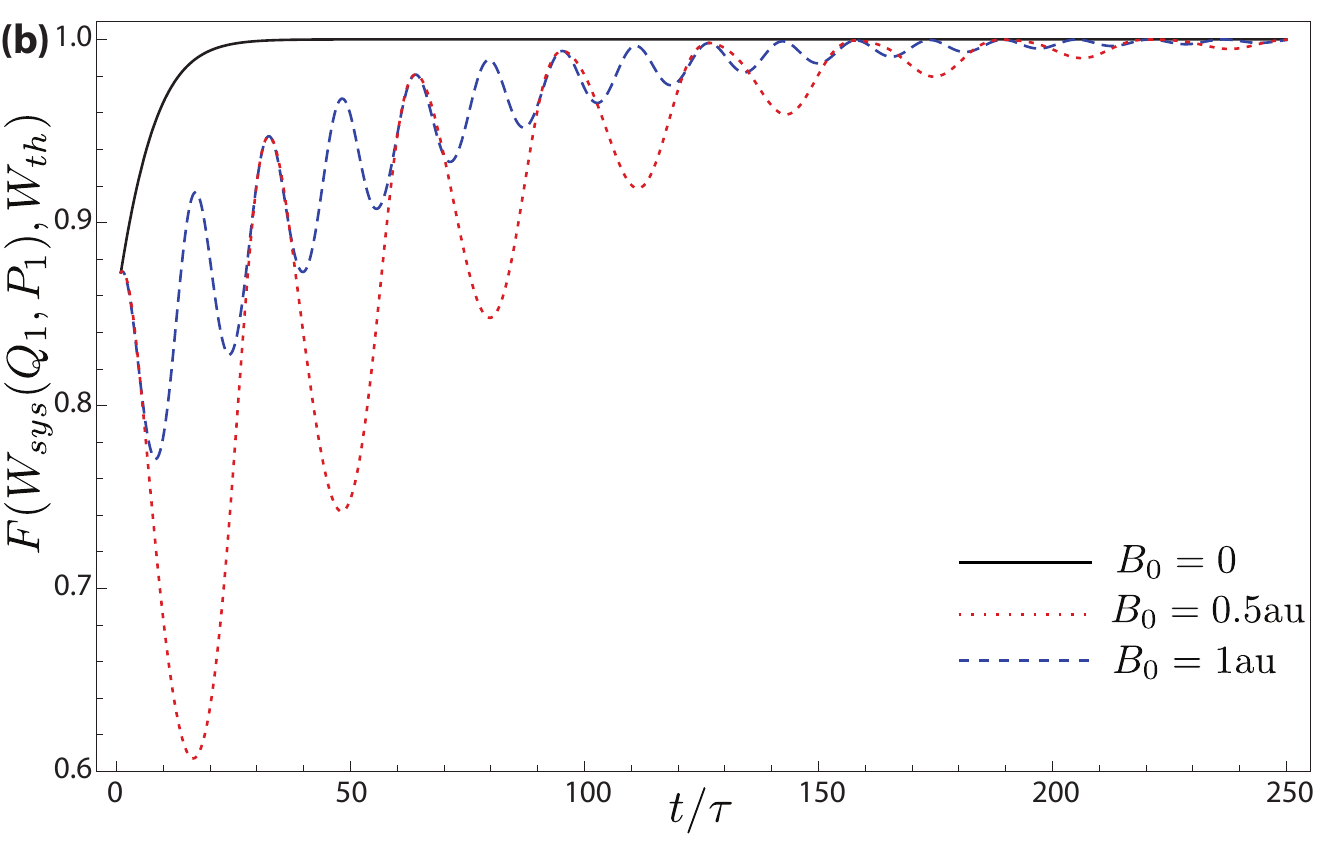}
\caption{(Color online) Quantum fidelity as a function of time for the Gaussian
state of the system $(Q_{1}P_{1})$ and the  asymptotic thermal state for complete cooling,
for a initial state of the system displaced from the origin, $(x_{0}, p_{x_{0}})=(1,1)$
( figure \textbf{(a)}), and for a initial state of the system initially in
the origin, $(x_{0}, p_{x_{0}})=(0,0)$ (figure \textbf{(b)}). The parameter $\tau$ is inserted
just to have a dimensionless variable; we choose a decay rate
$\Gamma = 0.1\tau$ and $\bar{n} = 4$ and $\bar{m} = 2$, such that we have the cooling
process.\label{figure01}}
\end{figure}

In Fig. \ref{figure01} we have plotted the quantum fidelity as a
function of time for the Gaussian state of the system $(Q_{1},P_{1})$
and the  asymptotic thermal state for complete cooling, assuming $(\bar{n},\bar{m})=(4,2)$
in order to have the cooling process of the system. We consider three
different cases, i. e., the absence of NC effects, $B_{0}=0$ (black,
solid line) and two cases with NC effects represented by $B_{0}=0.5$au
(red, dotted line) and $B_{0}=1$au (blue, dashed line) where au means
arbitrary unity. Moreover, two initially located system states were
assumed: $(x_{0},p_{x_{0}})=(1,1)$ (figure \textbf{(a)}) and $(x_{0},p_{x_{0}})=(0,0)$
(figure \textbf{(b)}). The black curves corresponds exactly to the process of cooling
of a Gaussian state when in contact with a thermal environment, i.
e., the fidelity increases monotonically up to unity. The only difference
is the initial value of the fidelity, which is easily explained by
noting that in the second case the system starts the dynamics with
$(\vec{d}_{sys})=(0,0)$, i. e., it is more closely to the asymptotic state for complete cooling process. For $B_{0}\neq0$, we observe that the monotonicity
of the fidelity is not fulfilled, though as the time increases the
state of the system goes to the asymptotic state.
It can be also noted a difference of phase relative to the two values
of $B_{0}$ when we change the initial position of the system in the
phase-space.

\subsection*{Non-Markovian-like Effect}

As we can note from the time evolution of the quantum fidelity in
Fig. \ref{figure01}, for $B_{0}\neq0$ the fidelity does not increase
monotonically up to unity. The presence of oscillations in the fidelity
during a thermalization process has been largely studied in the literature
as been associated to an information backflow from the thermal environment
to the system \cite{Rivas02,Breuer}. Non-Markovian effects is a current
field of research with several applications in quantum information
\cite{Latune,Luoma} and quantum thermodynamics \cite{Marcantoni,Maniscalco02},
both in theoretical and experimental areas \cite{Breuer03}. As we
have mentioned above, quantum fidelity obeys a significant property,
i. e., monotonicity \cite{Nielsen},
\begin{equation}
F(\Lambda\rho_{1},\Lambda\rho_{2})\geq F(\rho_{1},\rho_{2}),\label{ineque}
\end{equation}
where $\Lambda$ denotes a completely positive map, which serves as
a characteristic feature of Markovian dynamics (see Fig. \ref{figure01}).
Noticing that Markovian evolution guarantees a completely positive
trace preserving dynamical map $\Lambda(t)$, i. e., $\rho(0)\rightarrow\rho(t)=\Lambda(t)\rho(0)$,
which also forms a one-parameter semigroup obeying the composition
law \cite{kraus}: $\Lambda(t_{1})\Lambda(t_{2})=\Lambda(t_{1}+t_{2})$,
with $t_{1},t_{2}\geq0$. Therefore, any violation of the inequality
in Eq. (\ref{ineque}) is a clear signature of non-Markovian dynamics,
indicating that the associated dynamical map does not obey the composition
law. It is worth to mention that deviation from (\ref{ineque}) is
sufficient, though not necessary, reflection of non-Markovianity.
Following the standard procedure, we used the time derivative of the
quantum fidelity to indicate what we call non-Markovian-like effect,
i.e., an effective non-Markovian behavior due exclusively to $B_{0}\neq0$.
Therefore, for non-Markovian-like effect we have
\begin{equation}
\frac{d}{dt}F(\sigma_{1},\vec{d}_{1};\sigma_{2},\vec{d}_{2})<0.\label{derivative}
\end{equation}

\begin{figure}
\includegraphics[scale=0.5]{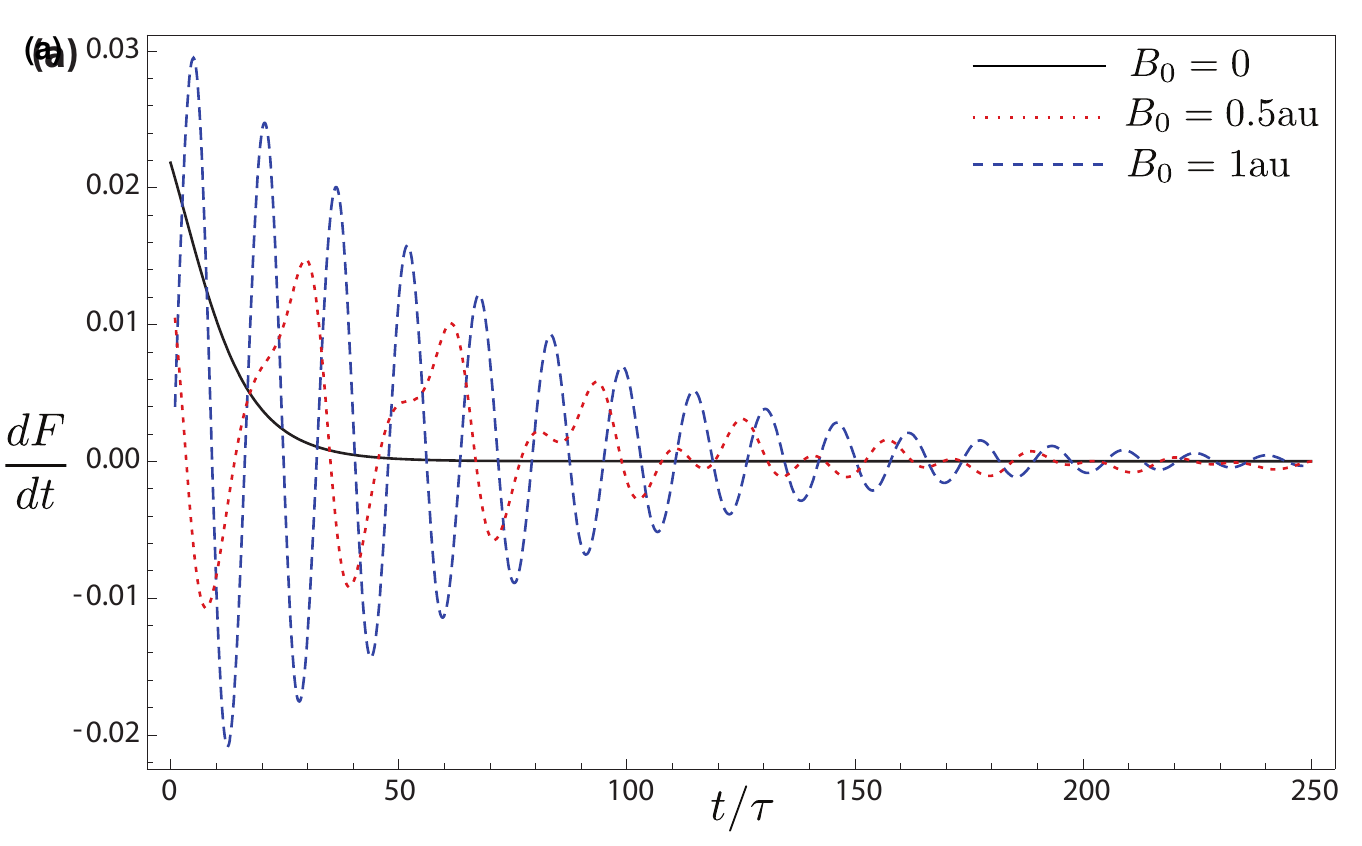}
\includegraphics[scale=0.5]{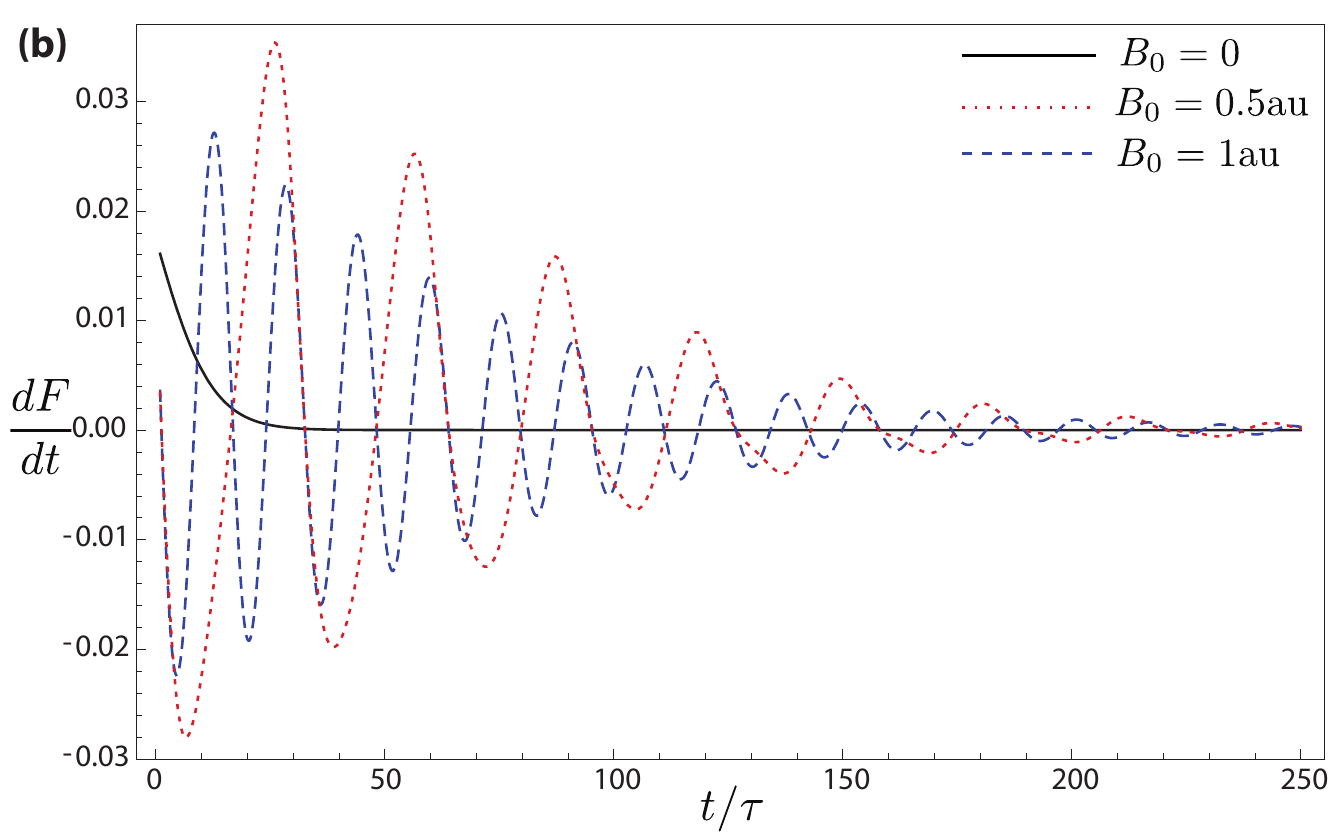}
\caption{ (Color online) Time derivative of the quantum fidelity for the same
cases represented in Fig. (\ref{figure01}). We considered three cases:
$B_{0}=0$ (black, solid line), $B_{0}=0.5$au (red, dotted line),
and $B_{0}=1$au (blue, dashed line). The same parameters of the Fig.
(\ref{figure01}) were used here. \label{fig2}}
\end{figure}

In order to visualize this type of effect in our system, we depicted
in Fig. \ref{fig2} the time evolution of the time derivative presented
in Fig. \ref{figure01}, using the same parameters. We note clearly
the existence of non-Markovian-like effects during the cooling process.
Obviously, the same effect would be present if the heating process
was considered, i. e., $\bar{n}<\bar{m}$.  From the Fig. \ref{fig2}
one can note that the intensity of the effect is greater as the intensity
of the field increases when the system starts out of the origin, and
the intensity of the effect is lower as the intensity of the field
decreases when the system starts at the origin.

It is important to worth that the thermal environment generates a Markovian dynamics on the system of interest. Then, any deviation from the Markovian profile described by the quantum fidelity must to have another source. Here, the only responsible source is the noncommutative effects mapped as effective external magnetic field.

\vspace{0.6cm}
\section{Conclusions\label{sec:Conclusions}}

In this work we used the possibility of mapping noncommutative effects
as external fields to study how these effects can influence a cooling
process of a system described by a Gaussian state. Using the well
known thermal diffusion approach, we investigated the time evolution
of the state of the system in two situations, first out of the origin
and second at the origin of the phase-space, remembering that the
asymptotic thermal state for complete cooling remains at the origin during the process. It can be observed that the evolution to the asymptotic state depends
on the the value of $B_{0}$ and on the initial localization of the
system in the phase-space. Besides, for $B_{0}\neq0$ the system takes
a longer time to reach the asymptotic state.

We also discuss the conceptual meaning of a non-Markovian-like effect
during the time evolution of the system from an initial state to the asymptotic thermal state. It must be stressed that the thermal
environment introduces a Markovian dynamics and that the decay rate
$\Gamma$ does not depend on time. Therefore, the negativity of the
time derivative of the quantum fidelity is exclusively due to $B_{0}\neq0$.
We believe that this work can contribute to unveil important features
of noncommutative effects mapped as external fields on the dynamics
of Gaussian states. Finally, in describing these effects through Gaussian
states and thermal diffusion, studies concerning the experimental
simulation of NC effects can emerge in the future.

\section*{Acknowledgment} Jonas F. G. Santos would like to thank CAPES (Brazil), Federal University of ABC, and São Paulo Research Foundation (FAPESP), grant 2019/04184-5 for support.

\end{document}